**A principled way to think about AI in education:**
guidance for educators and policy makers on
action based on goals, models of human learning, and use of technologies

Noah Finkelstein
University of Colorado, Boulder

The rapid emergence of generative artificial intelligence (AI) and related technologies has the potential to dramatically influence higher education, raising questions about the roles of institutions, educators, and students in a technology-rich future. While existing discourse often emphasizes either the promise and peril of AI or its immediate implementation, this paper advances a third path: a principled framework for guiding the use of AI in teaching and learning. Drawing on decades of scholarship in the learning sciences and uses of technology in education, I articulate a set of principles that connect broad educational goals to actionable practices. These principles clarify the respective roles of educators, learners, and technologies in shaping curricula, designing instruction, assessing learning, and cultivating community. The piece illustrates how a principled approach enables higher education to harness new tools while preserving its fundamental mission: advancing meaningful learning, supporting democratic societies, and preparing students for dynamic futures. Ultimately, this framework seeks to ensure that AI augments rather than displaces human capacities, aligning technology use with enduring educational values and goals. It is meant as a practical guide for anyone engaging in the use of new technology tools in their educational practices, or those people setting educational policies in the modern era.

*A new preamble (Jul 2026) and modest updates throughout the paper contextualize it since initial sharing (Jul 2025). The good news is that the arguments and principles of action posted originally appear to still be relevant one year later.*

**Preamble:**
A, or perhaps the, critical question that faces us with the regard to the use of generative artificial intelligence and related technologies (AI) is:

> In an era when computers and robots can engage in (or appear to humans to engage in) all human activity as effectively or more effectively than humans, what are those skills, habits of mind, and practices that we must maintain as within the human purview, in order to maintain our humanity and societies?

The question begins with an imagined future, and for the purposes of this paper and its associated recommended practices in education, we may not need to determine whether this possible future does or could exist.[1] No matter the answer, this future is worth considering, both because it is within the realm of possible futures and because such a future ought to inform our current practices. The question concludes by framing the goals -- or what is at stake currently -- for our humanity and society.[2] The concluding scenarios of this piece examine some short-term outcomes (e.g., a much more highly automated educational system); these are only steppingstones to the longer-term cultures we will build. And, finally, central to the question is what we as educators and policy makers have control over: defining what we do educationally - the skills, practices and habits of mind, directed toward the possible future we seek.

While the question is essential for our society as a whole to consider, we in education have a unique position to play in answering. As just noted, we have the opportunity to define and develop the practices, skills, and habits of mind of individuals in our society. In fact, this is why we constructed our educational system, and why it is so contested politically and in public discourse, today. We have the privilege and responsibility to act for the welfare of individuals and the public good, as will be discussed in more detail below.

While the role of the educator and the learner is examined in more detail below, three objectives for our educational practices stand out as essential. We need to focus on cultivating:
- <u>discernment</u>: the abilities to ask and frame appropriate questions, validate and contextualize answers to these questions, and to act appropriately with the results.
- <u>empathy:</u> the ability to understand and share the perspectives of others, which underpins our abilities to engage in communication and collaborative work
- <u>sense of self:</u> where students learn about their roles, purposes and whether and how they belong within a course, field, or broader cultural system.

These reiterate some of the thinking below, but are particularly important in the modern era. Accessing information, results of calculations, and analyses have become nearly instantaneous with modern technologies. Discernment develops not only abilities to engage in these practices, but essentially to ascertain the legitimacy, utility, applicability of the outcomes we get. It also informs what questions we ask, those that give rise to this information seeking behavior. Lastly, discernment shapes how we act based on the outcomes. Collectively, the practices of discernment support the development of metacognition (reflecting and regulating one's own thinking), and ethics, morals and judgement, as appropriate to the disciplines, communities, and society. Another key outcome and practice of education is our abilities to engage in collective work, and productive engagement in a cultural system (whether discipline or broader society); empathy underpins these capacities and thus, itself, is an essential practice and outcome for our educational system. Balancing our collective capacities is the development of self and sense of belonging. In our systems, learners can develop agency and purpose. These qualities underpin broader academic success, including performance, engagement, and retention within fields.

---

[1] Whether AI is simulacrum or engages in genuine forms of human activity, such as thought and empathy is an important question and long debated (e.g. Searle, 1980). Certainly, computers are different from people. However, the tasks, interactions, and forms of behavior they engage in can pass as human, and surpass human practices in many areas already. Their capacities, at the time of writing, continue to accelerate.

[2] Here, I am not seeking to be hyperbolic and consider the world turning into paper-clip factories (Bostrom 2003), but rather to argue that we have collective agency in defining what it means to be human and what societies are. Similarly, Pope Leo XIV has recently issued an encyclical examining these same matters (Pope Leo XIV, 2026).

Both the question and these answers are explored in more depth throughout this essay, providing key tactical choices we get to make as educators and policy makers. Here, I focus on the *what* - choices we individually make for our educational practices. As we seek to understand *how* to engage at scale, we must draw from models of institutional and organizational change in higher education – an inquiry beyond the scope of this piece. Notably, we have made enormous progress on this topic in recent decades (e.g., Henderson, et. al, 2011 and NAS, 2024). However, current models of institutional change generally presuppose our goals focus on *adoption* of proven, evidence-based approaches to educaiton. In the case of AI-infused educational practices, our evidence-base is much sparser, and yet they are *arriving* in our spaces anyhow. Hence, our models of educational transformation need to be updated. In a recent piece, we provide a framework and design implications for change initiatives to consider in the modern era (Perl-Nussbaum, 2026).

## Introduction:
## Why a Principled Approach to AI in Higher Education Teaching and Learning

> If generative AI and associated new technological tools can deliver instruction to our undergraduates in a just-in-time and student-specific manner (Kestin, et al 2025), why would our institutions of higher education continue (given their financial models and founding missions are based on undergraduate education)?
>
> If students can use generative AI to produce essays, homework solutions, and project designs without effort (Lohr, 2025), why would they spend the time learning (instead of more cheaply producing these products that are used to evaluate them)?
>
> If these new tools change how we think and conduct work in our academic disciplines (Kosmyna, 2025), how do we decide what we should do and what forms of thought are valuable?

As generative artificial intelligence (AI) and other technological innovations are becoming more widespread, are rapidly evolving, and likely to substantially change our educational system, it is worth developing key principles about the use of these tools to support learning and our learning environments. While a myriad of work has focused on (1) the promise and perils of these technologies, (Capraro, 2024) and (2) ways to implement these technologies in our learning environments (Mollick, 2025a, 2025b) here I seek to promote a third way - a foundational approach to the above – principles to help us think about the productive uses of new technologies in education. These principles build on decades of scholarship of teaching and learning and of technology use, and can inform how we use of these new powerful technologies to support our educational goals.

While considering both the largest scale vision (promise and peril) and on-the-ground practical steps (implementation) of technologies in education are necessary, a set of guiding principles can bridge the gap, connecting our implementation approaches to the large-scale goals we have for education. Many high-quality implementations and studies of AI in education are helping shape our use of these tools in the hopes of achieving broader goals. Each of these applications have principles that underpin their use; however, such principles are all-too-often implicit. That is, for example, when providing details on the use of generative AI as a tutor for students learning in college physics, or training students on prompt engineering for supporting the development of an argumentative essay, a variety of guiding principles underpin these pedagogical approaches, but often remain unstated (e.g., how are the tools used to support interactive engagement, balance cognitive work between human and computer, etc.).

Not only do principles of use allow us bridge between high level goals and on the ground action, but they also give us tools to ethically adapt, experiment, and implement as these tools evolve. We are at the beginning of a grand disruption in our fields; committing to principled-based choices and actions that are based on decades of scholarship will allow us to bootstrap, invent, learn, and revise our approaches while maintaining the promises of a relevant, moral, and effective educational system. By more thoroughly articulating guiding principles, we can: seat technology-based educational practices in the broader constellation of educational activities; ensure that these new approaches align with, and in fact accelerate our capacities to realize, rather than undermine our goals; create adaptable, ethical approaches that evolve with the technologies; and develop a robust strategy for the sustained and successful implementation of AI and other technologies.

These technologies are particularly rapid in its evolution and deployment. And, much like AI itself, where AI is headed is hard to predict. To that end, while it is hard to make predictions, especially about the future (Steincke, 1948), we might do well to skate to where the puck is going to be rather than where it is or has been (Gretzky, W., n.d.). We have grand agency and capacities for shaping the future of education, And, if we do not proactively and engage in the use of these new technologies for education, they will still be present. They will still be used. However, they are more likely to use us - to shape our interactions with each other and our students that we did not intend, than if we had sought to intentionally engage. With appropriate forethought, these new tools can help us advance our capacities and the capacities of learners to realize the future as it ought to be.

In seeking to productively develop and engage with these new tools in our educational environments, it is worth considering some basic questions:

- What are our goals in education?

- Given practical and actionable definitions of *education* and of *technology*, how do we consider appropriate roles for educators, learners, and technologies?
- How will the use of these technologies transform what we are learning?

This essay seeks to explore these questions to provide the reader a set of principles by which they may bridge between implementation strategies and realize the promises and avoid the perils of these new technologies in education. These principles offer guidance for those seeking to understand and deploy new technologies in their educational environments, both formal and informal. Because our educational contexts vary dramatically (and in fact are a strength and hallmark of the American education system), these principles are designed to be flexible and help allow us to contextualize the use of these technologies to the specific environments educators find themselves in, as well as to allow for a variety of goals of education. The present focus is on colleges and universities; though, many of the approaches will likely apply to pre- and post-college environments.

**Why educate?**

While there are many excellent volumes written about the purpose of education (e.g. Dewey, 1916), starting with such a fundamental question is essential in ensuring that our educational practices are aligned with our goals. Three broad classes of goals span the space of why we have constructed our educational systems:

1) Developing individuals. Learning is the mechanism by which people develop (Vygotsky 1978). The development of higher order cognitive functions not only supports basic skills for engaging in society (language and mathematical literacies, reasoning, argumentation, communication, socialization, empathy, agency, identities, and more), but also makes individual's life more meaningful, richer, healthier, and connected.
2) Societal infrastructure. The roots of most modern societies are grounded in an educated and informed populace. Democracies are dependent upon basic skills and common culture developed through education. As our societies become more interconnected (both within and across communities), as the decisions that one collective makes increasingly impact other collectives, education provides capacities to understand and address these interconnections in ways that are to the benefit of all.
3) Workforce development. The most common, modern framing of the purpose of education is to get a job, or more appropriately, to support a career. Arguments around workforce development vary from specific skills training and certification to gaining broader skill sets and supporting the capacities for job and place-based learning. Motivations also vary from fulfilling the needs of industry and the economy to fulfilling the interests and needs of our learners.

Principle 0: Know why you are acting -- whether at the highest level or specific class activity. What are the goals you are seeking to advance? Ultimately the goals one prioritizes ought to inform the actions one takes and how one may productively leverage new technologies. These goals can be multifaceted, and potentially contradictory, as happens regularly in our education systems.

The list of goals above is placed in (my) priority order. While workforce development and job training goals are the most commonly cited, they are achieved by addressing the initial two goals - developing learners and preparing them to engage in society (AAC&U, 2009; Finley, 2021).

Developing basic reasoning and communication skills and understanding historical and humanistic traditions provide students the skills useful for today's and tomorrow's jobs. However, preparing students for specific workforce needs and associated skills, does not necessarily prepare them as active and engaged citizens, nor to develop their own perspectives and passions. Others may have other goals for our educational system. No matter the objectives, being explicit about our goals ought to guide the use of technologies in creating our educational environments. Will technologies support the development of more self-realized, empathetic, community engaged and connected, rational, civic, and artistic individuals?

**What is learning? and teaching?**

Again, volumes have been written about what learning is and how and where it happens. And while there are perfectly valid definitions of *learning* based on biology and brain changes, e.g. learning is the creation of neural pathways, such definitions are less operational than complementary perspectives drawing from socio-cultural theories of learning. Learning is about socializing humans into cultural systems (Vygotsky1978 / Cole 1996).

> Learning is the internalization of culturally developed knowledge and practices. It occurs through the interaction of humans with other humans directly and indirectly, through the use of tools.

Thus, culture shapes learning and vice versa. Whether a child learning arithmetic - internalizing what 6x 4 = 24 means (e.g. six groupings of four objects) along with the associated algorithms, or what the four freedoms enshrined in the First Amendment of the U.S. Constitution are and how they apply today, or how computer-aided-design software may be use in the manufacture of assistive technologies, people learn material deemed important by cultures through interactions with other more knowledgeable peers, culturally relevant tools, and their proxies.

An appropriate parallel consideration is where learning ought to occur – i.e. in formal settings (schools / colleges), on the job training, or informal and community settings. However, such considerations are beyond the scope of this piece, which will focus on institutions of higher education and assume their value and relevance (for the moment).

**What are these advanced technologies?**

When thinking about technology it is easy to immediately think of new "hi-tech" tools and silicon-based machines. And, while these indeed are technologies, it is productive to broaden the lens so that we consider other technologies and humans' long history of developing and using these technologies -- for example, pre-silicon-based technologies include: a hoe, a pencil, language, and or our disciplinary domains, such as physics. Notably these technologies (or tools) are both material (a hoe) and intellectual (physics). A functional working definition is useful in considering technologies in education (Cole, 1996; Farrell, 2025a, 2025b):

> Technologies are human constructed tools, both material and intellectual, that reorganize how humans interact with each other, with the world, and with other technologies.

Consider a pencil and paper – these technologies modify both human memory (expanding our capacities to track a larger number of data bits than our short-term memory tracks alone) and our abilities to communicate (expanding communication across space and time). Of course, Socrates has a point that writing may come at the detriment of our internal memory systems and possibly oral communications. Similarly, language systems or disciplinary fields change how humans interact. And of course, generative AI, LLMs, machine learning, and other rapidly evolving computer-based tools reorganize human practices and capacities.[3]

**Deploying these new technologies for education**.

Drawing on these actionable definitions of *learning* and *technology*, we might consider how humans use these new tools to productively support our *goals* of education. A natural response is to consider: what are these new technologies good at compared to what are humans good at, and subsequently allocate responsibilities based on these capacities. For example, until the early 2000s computers, which had been around for decades, were very good at computation and poor at vision (parsing images). As such, much of computer work had replaced manual computation by humans and humans remained in charge of discernment in image production and detection. In recent years that has changed, and notably computers have become very capable of image detection and processing - often and initially with significant errors. But this scenario raises the questions:

> Just because a technology is capable of a task, should it be assigned to and responsible for that task?
> If new systems can teach, should we replace educators?
> If these technologies can learn, do they replace students?

Of course, these are hyperbolic and provocative framings. Rather than such an absolutist and reductionist framing, a potentially more productive approach will be to consider how these new and emerging technologies might advance human capacities.

> Principle 1: In our educational enterprise consider what domains of work are necessary or essential for educators to maintain as human led and directed, and what is feasible and safe to outsource to these new technologies.

> Principle 2: Determine what practices and activities are necessary or essential for students to engage in as human led and directed, and how newly advancing technologies might advance learners' capacities.

Notably, these principles do not prescribe binary actions. There are degrees of responsibility to be allocated, and simultaneously, new responsibilities will emerge, because of the development of technologies. To that end, we ought to consider what are the current and near-term roles and responsibilities for educators and learners.

***Roles of the educator***

---

[3] Technologies are typically neither good nor bad- it depends upon how they are used. Though, notably, technologies do come with predispositions for use, which embed values. Pope Leo XIV has an excellent examination of this point in his encyclical on AI (Pope Leo XIV, 2026).

There are a myriad of roles that educators serve in advancing learning, and, yet again, volumes have been written about the roles of teachers (NAS 2025). Here, zooming in one layer further from the goals of education that we opened with (individual development, societal infrastructure, and workforce needs), we might consider what are the classes of activities that an educator (in higher education) engages in and has control over. Here, we consider them from a design or reverse engineering perspective. Knowing what our goals are, how do we achieve these ends? And, how do we know if we have achieved them?

The following are key classes of action that an educator engages in – they ought to shape and will be shaped by the uses of technology. These roles include:

**Establishing the objectives** of a course, units, and lessons. While not established in isolation, it is the instructor who contributes to and enacts activities with various goals, both explicit and implicit, in our classrooms.

Content, Practices and Skills development. These are the dominant goals, or learning objectives, in our classes. Does a student know what Newton's Laws are, and when and how to apply them by developing algorithmic proficiency? This may also include defining the bounds of a domain, e.g., what are and are not considered "physics questions".

Ways of thinking, habits of mind, and metacognition. These are the tools that we use in developing and applying our content understanding and skills. We educate students in the habits of mind of a field, the appropriate intellectual moves, and methodologies. Simultaneously, we teach students to think about and regulate their own thinking – metacognition. Finally, these habits of mind may include discernment, ethics, and morals within a field or cultural system, or society more broadly.

Motivation, empathy, trust, identity, purpose, and community. The educators and community of learners in a course contribute to individuals' commitments - personal and collective engagement in the field. Within a course, educators cultivate participation (or not), build a sense of trust, belonging, and identity of individuals within the field (or not), and make the field relevant for individuals and broader communities, frame the purpose of learning and the domain, and build a sense of community within the class and domain. Many of these characteristics are underpinned by and developed through empathy, which can be cultivated (or curtailed) in our classes.

Principle 1.1: identify which goals your classes have, and make these goals explicit in educational practices, through the use of modern technologies (including generative AI). Note which goals may be subverted through the use of generative AI.

For example, is learning syntax of a programming language essential or outmoded? And how might the use of new technologies support or circumvent goals for deeper reflective thought in computational problem solving, or cultivation of more members of our communities engaging in computational thinking? Is writing an essay "independently" necessary? And if not, how do we ensure the students develop attention and emphasize clarity of thought?

**Curating information, knowledge, and resources** of a domain. An instructor curates learners' access to information, knowledge and disciplinary tools and practices that are in service to the goals of a course.

Information. Historically, education occurred in an information-limited age. An educator provided access to information. A Personal library (having access to information) was the sign of an educated individual. In relatively recent decades, we moved to (more or less) ubiquitous access to information. The role of educators has, in fact, now been shifted to limit, vet, and validate the information students attend to, to assist them in focusing their attention on relevant information.

Knowledge. The organization of information, or knowledge, is one of the desired outcomes of education, and a form of content in a domain. An educator curates knowledge, organizing the information that students have access to, so that these knowledge structures allow a student to productively integrate and assimilate additional information and develop insights.

Resources. An educator provides access to essential material and intellectual tools and governing practices around these resources. Whether laboratory equipment or library search tools, a calculator, or perhaps generative AI engines, an instructor introduces, curates and guides learner's use of domain-appropriate tools and practices, and their application.

Principle 1.2: Consider how the curation of information, knowledge and resources may be augmented and needs to be adapted in light of new technologies, such as generative AI tools.

For example, and perhaps most obvious, how should generative AI be used within a domain and how should students be taught to use it? If new technologies begin curating information, and indeed knowledge structures, how can instructors utilize these resources productively? I would love for generative AI to replace the clicking I currently do in creating and updating my Orwellian-named learning management system. Commensurately, how should students be taught to use (and validate) technology-based information, knowledge, and resources, not only for this course but as a foundational life-skill?

**Design of the educational activities** (and possibly the environment) appropriate to the learners who arrive. A great deal of scholarly work in recent years has attended to pedagogical approaches and curricula for learning -- both within and outside our classroom environments. Key principles in the design of educational activities are nicely summarized in a recent consensus study from the National Academies (NAS, 2025).

Actively engage students. Decades of evidence and foundational theories of learning note that learning is an active process (not the absorption of information), and environments that proactively engage students show demonstratively higher learning gains than our traditional, one-way, lecture environments.

Leverage students' backgrounds to motivate and engage. Students' varied interests, goals, knowledge, and experiences impact what and how students learn. Support the affective, social, identity, and community development of learners. Whether a student engages or not, whether they learn or not, will depend upon how they feel about the domain, their identities, and sense of belonging.

Build adaptable and transparent environments. Sharing the goals, processes and outcomes with learners and the communities will ensure more students are more successful. Adapting to student interests and needs is essential if education is to be seen as relevant and necessary.

Principle 1.3: Consider how technologies may assist in each of these design components of effective instruction. What are the benefits and liabilities of replacing or augmenting instructor activities with AI? How might technologies assist in the selection of effective, educational activities that are responsive to the students in a particular course?

For example, many basic, skills-based activities may be served by a computer tutor (think typing-tutor from a certain era), and while it might be tempting to personalize all interactions for a given student (Stephenson,1995), such activities may undermine a goal of socializing a student and helping them regulate their own learning, abilities to interact with others, and navigate uncertain and complex situations. In a complementary approach, it may also be the case that AI can serve as a concierge of sorts for faculty in designing their educational environments.

**Assessment and Certification.** A core purpose of education is assessment and certification of individuals' educational outcomes and evaluation of the educational approaches themselves.

Assessment of past, current and future states of learners. To be effective, educational activities need to meet students where they are, building on what they already know -- hence, assessing the prior state of student learning is key. Historically, this has occurred through course sequencing and prerequisites. More modern (and expensive) approaches include assessment of students on entry to a program or class. Assessment of the current state of the learner can be developmental (formative assessment) both for the student and of educational practices themselves. Often, this formative assessment occurs through lower-stakes assignments, such as homeworks and quizzes. Assessment of the end-state of learners in a class, most often represented by student performance on exams, essays and other activities, and ultimately a course grade, is the summative form of evaluating students. In some instances, and potentially very valuable, we can measure the future trajectories of students, documenting what they will know and be able to do in the future (Vygotsky).

Certification. Ultimately instructors certify the understanding of students enrolled in their courses. They certify that a student is ready to continue, or not, in a course sequence. Students, future educators, potential employers and others in society have vested interest in certification of individuals, especially as this certification captures student understanding, skills, practices, and the goals presented above.

Principle 1.4: Consider in what ways technologies may assist in the assessment and certification of individuals. In what ways do humans need to exercise professional judgment and to what degree might this be outsourced to a machine?

For example, our current summative assessment and certification is quite coarse, symbolized in a grade, certificate, or degree. It represents aggregate assessment of skills, knowledge, ways of thinking, and proficiencies. That is, grades are not invertible functions.[4] Technologies may be able to provide a compact accessible representation of student capabilities in given circumstances that may be unpacked, or delved into, as it suits the interests of the learner and others seeking to work with the learner. Many novel efforts are seeking to expand and tailor our capacities as educators to provide individualized actionable feedback, at scale for our classes. These features may become particularly important roles as we develop more certificates, badges and micro-credentials.

Assessment practices and continuous improvement: Based on the educational outcomes of learners, and other considerations (such as resources required and institutional constraints), an educator ought to engage in a process of continuous improvement informed by the specific context, student outcomes (and student input), and by scholarship in the field (NAS 2025).

Principle 1.5: Consider how to leverage technologies to inform curricular and pedagogical adaptations based on student assessment.

For example, an instructor might leverage generative AI to curate scholarly and evidence-based recommendations for educational practices to address a given situation. And, there may be an opportunity for technologies to analyze the breadth of student performance data in a course to provide summaries and potentially, even recommendations for tailored pedagogical approaches.

***Supporting Learners' Roles***

Students' roles may parallel and complement those of the educator, listed above. Glibly, if the role of an educator is to teach, the role of a student is to learn. If an instructor dominantly defines the goals of learning, students can simultaneously contribute to these goals and contextualize them for themselves. If an educator is to curate information, knowledge and tools, the student ought to access, use, and internalize these. If an educator is to actively engage students, the students are to engage. And so forth. The following takes a learner-centered perspective and can help inform actions for the student, instructor, and educational leaders.

**Attending to goals:** What are the necessary approaches a student ought to engage in to attend to the highest-level goals listed in the introduction? What are the student goals for participation in an activity, course, or degree program? If a student is seeking a job or certification, their focus may be on content, and associated skills, for example, knowing how to engage in computer aided design, coding or developing technical writing skills. If a student is seeking a career (multiple jobs) and to join a community, their focus may additionally include understanding ways of thinking, discourse patterns, norms, and habits of mind of the community. If a student is seeking to define a community, their focus may additionally include attention to developing metacognitive skills, discernment, ethics, morals and purpose. And personal development of an individual may additionally involve the attention to identity development, motivation and purpose.

[4] Many thanks to Edmond Johnsen, CU Boulder Center for STEM Learning, for this framing.

Principle 2.1: Ascertain, attend to, and develop the goals our learners. What are appropriate and needed goals in our future, given the new capacities and trajectory of technologies?

For example: if a student arrives with the intention of only learning specific skills, we can provide opportunities to develop those skills, potentially utilizing new and emerging technologies, such as computer-tutors. At the same time, we might help make students aware of broader opportunities of education and engage them in understanding the future roles of technologies in their lives.

**Practicing competencies and acting on assessments:** Each of these goals will have related practices, skills, and competencies that a student may develop. A student will need to engage in routine practice, with feedback to develop the associated content and skills. How does a student respond to the feedback provided (whether formative and summative)?

Principle 2.2: Carefully consider which practices students are engaged in (both those they are asked to engage in and those practices beyond what they are asked to engage in), and ascertain in which ways generative AI and new technologies may assist and may circumvent the goals that the students and the instructor hold.

For example: A great deal of recent attention has been paid to students' use of generative AI in computer coding classes (Lohr, 2025). In many courses, some students are producing code through the use of AI tools, rather than producing code themselves, as required by the course. Here, some of the challenge is that student performance is not aligned with learning, and may not be aligned with instructor goals. Similar challenges show up in the humanities (O'Rourke, 2025). Can we support students' clarity of thought and expression, develop their values and voice, and minimize the temptation to outsource thinking and attention? How might we adapt our educational activities to address these challenges? Do we regulate students (holding in-class, supervised assessments), or do we modify the goals and activities that students engage in to prepare them for a new technology-laden landscape?

**Applying and Synthesizing:** In our current system, much of our educational practice relies on the student to synthesize across courses, to contextualize and apply this understanding to their own lives and trajectories. Of course, there are some courses that look to span disciplinary content, e.g., capstone classes and internships provide some structured opportunities for learners; however, these are far from the norm.

Principle 2.3 Consider the ways new technologies might assist or circumvent the opportunities to analyze and synthesize across courses and disciplines.

Example: There is some movement to have student generated portfolios, using new technologies. Such portfolios can span across courses and serve to complement student grades and instructor and institutional certification, demonstrating what students can do. Of course, the role of authorship – to what degree is a portfolio student versus AI generated – will be a matter for all to consider.

**Collaboration, including co-design of educational environments.** Whether implicitly or explicitly, our educational environments teach students how to collaborate and contribute collectively (or not). And, students play an essential role in the design and enactment of our classes themselves. Most often students' roles in the co-design of educational experiences are implicit. That is, students can vote with their feet, participating in various elements of a course, and in practice, to grow or shrink demand. Increasingly, while attending to goals, educators are spending some course time working with students to explicitly design the educational experience, to ensure that it is relevant, meaningful, and impactful. Whether establishing collective course norms, for example around cheating or use of AI, or around content, allowing students to define, to some degree, what is covered in a course and how it is covered, students can partner with and make more effective learning environments.

> Principle: 2.4 Review the roles that students might serve in collaborative design and enactment of our educational environments. Instructors and others can work with students to consider how generative AI might advance the goals and support the practices (such as those established in 2.2) in our classes.
>
> For example, we can consider working with students to define what roles generative AI and other novel technologies might serve in a class. Can students co-generate a set of policies and governance practices in class round the use of AI as part of class activities? Additionally, an instructor might work with students to identify educational practices that prevent the nefarious uses of generative AI, e.g. those that circumvent the learning goals, and instead proactively align performance measures with learning and understanding.

**Review our educators / system.** Students, dominantly through rating systems, evaluate the instructor and course for institutional and individual feedback. These reviews occur through formal college-sponsored mechanisms, third party sites, or word of mouth. They also implicitly rate a system based on their participation - do they enroll? As our institutions become more dependent upon student tuition as a funding source, such explicit and implicit ratings will only increase in value. Notably, however, our current rating systems suffer from not being coupled (correlated) with the instructor-defined goals of our course, nor to measures of faculty effective practices, nor to the highest-level objectives and goals listed at the beginning of this piece.

> Principle 2.5: Consider the roles that new technologies might serve in supporting students' evaluations, as well he collection of evidence and their analyses used for evaluation.
>
> For example, building on longstanding adaptive technology use, these technologies can be useful in the collection and rapid analysis of weekly feedback that the instructor may use to inform their course in real time. Here, technology may be particularly useful in large enrollment courses. However, technologies may also be useful in the large-scale, detailed analysis of student work (not for grading) but for understanding common student difficulties around particular content. Technologies may also serve to ascertain a given instructor's development over time, measuring faculty teaching effectiveness, in part, based on student learning and development over time.

Are these the roles we want or need for our instructors and students? As we rethink our educational system, whether to modify our courses modestly to incorporate new technologies, or grandly to reconsider the structure and objectives of higher education itself, our framing of goals, roles, and the associated principles can be useful tools.

***Administrative roles***
While the focus of this piece is on those individuals enacting and engaging in educational activities, administrators and staff who shape and support those directly involved serve key roles, as well. These administrative roles include department chairs, deans, those in centers for teaching and learning and technologies offices, as well as those in higher level leadership such as the provost and academic leadership cabinet. In addition to considering the principles above and how these may be supported through administrative actions, these leaders can provide guidance, resources, and level-setting of expectations. Administrators ought to clearly communicate an institution's commitment to and policies around these emerging and rapidly evolving technologies. These statements include but are not limited to clarifying what constitutes (and what violates) appropriate use of technology to support campus principles of teaching, learning, and engagement, and how these practices are evaluated and rewarded. Resourcing the adoption and effective implementation of these tools will also be necessary. Resources need to include material support (funding and time) and intellectual (professional development and recognition of work). Creating faculty learning communities, communities of practice and transformation, and materials for sharing can advance the capacity of individuals and the institution. Finally, administrators set expectations both of those enacting and using these tools (faculty and students), and of the administrations' expectations of the new workload and practices of instructors (and students). We ought to take a lesson from COVID-era education where everyone was asked to do more, simply adding to the workload, which led to faculty overwork and burnout. Administrators need to consider what to take off educator's plates, and potentially how new technologies may ease rather than further add to workload.

## Scenarios

A few scenarios provide the application of multiple principles at once to current and potential coming educational circumstances.

Ubiquitous and university-supported access to generative AI tools: Institutions are now providing access to subscription service use of AI as part of student enrollment. One can imagine a world where these new technologies are made available and promoted, but do not come with requisite preparation, support, or guardrails. Without commensurate training for both students and educators on how to use these tools effectively, we will miss this moment and may make things worse. Faculty may ignore these new tools and students use them in unsupervised fashion, simply plugging essay prompts or homework questions into an AI engine and reporting the results. In this approach, the enacted goals may devolve to the lowest level, performance vs. deeper learning, and not developing students' metacognitive, reflective, and higher order skills (1.1), and students' goals (2.1) remain undeveloped.

Alternatively, faculty may seek to regulate technology use in a course, going "back to basics" and having students take in-class exams and writing essays in blue books, which sends messages that may counter our overt goals -- e.g. suggesting that professional practice in our disciplines happens in a 50 to 75 min block, where answers are known in advance, and individuals have to work in isolation and without the resources available more generally. Thus, AI may not reframe class but serve to limit current practices (1.3), and undermine the faculty shaping of class (1.1, 1.2). Cheating (2.2, 2.3) may increase, if practices (1.3, 1.4) and goals (1.1) do not adapt. Of course, there will be individual instructors leaning in and working with students on new class designs (2.4, 1.5) but if this is not supported and scaled across campus (as per the description of administrative roles above) it will remain an uneven patchwork. This suite of principles may be useful in creating stepwise, wide-spread engagement across campus to share collective new models and approaches.

Supporting basic skills development (helping students "catch up"). An increasing fraction of students arrive at college without foundational mathematical or writing skills; COVID exacerbated these preexisting trends. Many have long-considered technology as a mechanism to support the development of these students. Indeed, there are examples of technology enhanced skills development (ALEKS, 2025; Carnegie 2025) which go back to the early days of the typing tutor. Arguably these approaches can support more advanced goals in a course (1.1., 2.1), utilizing our limited class time on practices that support these goals (1.3,1.4, 2.2, 2.3) and even afford the time to engage with students on class design (2.4, 1.5). However, we must be extraordinarily cautious and careful in such approaches. It is easy to falsely attribute a student poor performance, e.g. on a mathematics placement exam, to student capacity or preparation. In fact, poor performance is often coupled to lack of preparation and support in learning how to learn, motivation, trust in education, poverty, or other life circumstances that would not be addressed by a technology tutor. These students may particularly benefit from enhanced interaction with other caring humans -- not necessarily technologies. Without attention, educational innovations tend to exacerbate rather than reduce learning gaps between well-off and lower-resourced individuals. (Reich, 2020).

Training the AI, developing professional practices. If, in order to learn the basics of a field, including, but beyond the traditional content, what if we had students developing and training AIs in the subject domains they are learning? Given that LLMs are customizable, one could imagine coaching students on how to develop their own agents for further exploring a content area. Additionally or alternatively, given their capacities for coding and rendering, AIs can be leveraged by students to represent a content field or domain. Recently we've had students (who have no coding experience) designing computer simulations to represent basic to advance physics phenomena (BenZion, 2025). In so doing, students must validate AI based approaches both in terms of the outcomes and scientific accuracy (1.4, 2.4, 2.5). Of course, careful guidance and scaffolding is necessary, but this approach can support the skills (1.3) that are part of our professional domains (1.1, 2,1) including: communication, discernment, metacognition, and trust. In parallel students learn about the strengths and limitations of these new technologies (2.5).

Addressing Teaching Assistant (TA) and staff shortages. It is entirely likely that there will be increased teaching demands placed on our faculty and potentially fewer support resources (TAs), especially at research-based universities. As an alternative to simply increasing number of courses assigned to individual instructors, we might consider novel approaches through the use of new technologies. What tasks may be automated? Hopefully, some of the course management roles (producing a website, attending to the learning management system, etc.) can largely be handed off to AI-agents, leaving the tasks that require more judgement to instructors. Faculty and students can focus on goals and defining practices (1.1, 1.3, 2.1, 2.5), but technologies may be able to give faculty feedback on modern sources for curation (1.2), for example providing curated lists of videos to complement (or replace?) our textbooks. It also appears that technologies may be able to provide feedback to students on written work (2.3, 2.5), based on an instructor defined rubric (1.4). In this case, an instructor (or TA) could evaluate student performance based on how they used, responded to, and justified their responses to AI-produced feedback on work, whether a homework set or essay (1.4). In such a scenario, in addition to engaging in current advanced skills of a domain (2.1, 2.2), students would be learning about effective use of AI and potentially helping shape how the domain uses AI (2.3, 2.4).

**Addressing the opening questions:**

> *If generative AI and associated new technological tools can deliver instruction to our, why would our institutions of higher education continue?*

In so far as we consider our institutions only about content delivery (access to information) or about basic skills acquisition (e.g. certain forms of computation), we will not need our institutions of higher education. However, so long as we hold the goals of advancing the lives of individuals and building a broader society for the public good, then we will need the universities and the models they hold for educating individuals through socialization and development. Through our institutions of higher education we develop learners, build communities and connections, and cultivate habits of mind necessary for a functioning society.

> *If students can use generative AI to produce essays, homework solutions, and project designs without effort, why would they spend the time learning?*

Humans are born and evolved to learn; we are learning machines. However, sometimes in our formal educational systems, we have too strongly decoupled performance measures from learning practices. The principles above help us recenter our practices, to partner with learners, to create environments where authentic activities and learning are enacted, valued, and rewarded. Realigning our metrics and practices with learning will reinforce and reward desired forms of student engagement.

> *If these new tools change how we think, how do we decide what we should do and what forms of thought are valuable?*

Human higher order cognitive functions have always been modified by the tools we use -- whether language or writing. These new tools will change our cognitive processes. How these change us, in part, will be up to those framing the roles of AI in education (and society more generally). Leaning in to and understanding key roles for humans to play – in teaching and learning – and the broader goals – why we are educating– can give us sign posts to a more empowered, capable, successful, and technology-enabled future.

**Conclusion:**
Whether promise or peril, or which practices we engage in, we are at the early stages of a grand experiment around the use of next generation technologies that will likely radically reshape education. A guiding set of principles, based on the thousands of years of how humans learn and how we have used technologies can assist in understanding what we are doing as well as nudge us in directions we seek -- improved opportunities for individuals and society alike. This principled approach to consider how we use of generative artificial intelligence and other technologies serves as a bridge between practice-based approaches with the grandest visions of its use. By ascertaining and applying evergreen principles of learning and technology use, which I begin to enumerate above, we can shape how modern technologies are deployed in service of our learners, the education system, and society.

Acknowledgements:
I am most grateful for the thoughtful feedback and improvements to these ideas made by: Howard Gobstein, Peter Grunwald, Michael Lightner, Diane Sieber, and Phoebe Young